\documentclass[12pt]{article}
\usepackage{amssymb}
\usepackage{amsmath}
\usepackage{color}
\usepackage[unicode,bookmarks,bookmarksopen,bookmarksopenlevel=2,colorlinks,linkcolor=blue,citecolor=green]{hyperref}

\input{tcilatex}
\begin{document}

\title{\protect\vspace{-1cm}\textbf{Oriented Associativity Equations and Symmetry Consistent
Conjugate Curvilinear Coordinate Nets}}
\author{Maxim~V.~Pavlov$^{1}$, Artur Sergyeyev$^{2}$ \\
$^{1}$Department of Mathematical Physics,\\
Lebedev Physical Institute of Russian Academy of Sciences,\\
Leninskij Prospekt, 53, Moscow, Russia;\\
$^{2}$Mathematical Institute,
Silesian University in Opava,\\ Na Rybn\'{\i}\v{c}ku 1,
74601 Opava, Czech Republic\\
e-mail: maxim@math.sinica.edu.tw, Artur.Sergyeyev@math.slu.cz}
\date{}
\maketitle

\begin{abstract}
This paper is devoted to description of the relationship among oriented
associativity equations, symmetry consistent conjugate curvilinear
coordinate nets, and the widest associated class of semi-Hamiltonian 
hydrodynamic-type systems.
\end{abstract}

\begin{flushright}\textit{In honour of Franco Magri}
\end{flushright}

\tableofcontents

\bigskip

Keywords: Riemann invariants, Haantjes tensor, hydrodynamic-type systems, oriented associativity equations, centroaffine geometry


\section{Introduction}

The nonlinear partial differential system\footnote{%
Here and below the sum over \textit{any} of repeated indices in the opposite
locations (that is, one subscript and one superscript) is understood;
otherwise the sum is indicated explicitly.}%
\begin{equation}
\frac{\partial ^{2}c^{i}}{\partial a^{j}\partial a^{m}}\frac{\partial
^{2}c^{m}}{\partial a^{k}\partial a^{n}}=\frac{\partial ^{2}c^{i}}{\partial
a^{k}\partial a^{m}}\frac{\partial ^{2}c^{m}}{\partial a^{j}\partial a^{n}}
\label{ori}
\end{equation}%
describing a displacement vector appears in \cite{Dubr}. Following \cite%
{LosevManin} we call this system the \textit{\textbf{oriented associativity
equations}}. This system admits the scalar linear spectral problem (cf. \cite%
{Dubr})%
\begin{equation}
\frac{\partial ^{2}h}{\partial a^{i}\partial a^{j}}=\lambda \frac{\partial
^{2}c^{m}}{\partial a^{i}\partial a^{j}}\frac{\partial h}{\partial a^{m}}
\label{zak}
\end{equation}%
or, alternatively, the vector linear spectral problem (cf.\ e.g.\
\cite{Art} and references therein)%
\begin{equation}
\frac{\partial b^{i}}{\partial a^{k}}=\lambda \frac{\partial ^{2}c^{i}}{%
\partial a^{k}\partial a^{m}}b^{m}.  \label{flow}
\end{equation}

Integrable system (\ref{ori}) was extensively investigated in a number of
papers (see, for instance, \cite{KonopMagri}) dedicated to the so-called
coisotropic deformations. Some other aspects were considered in \cite{HM},
\cite{LP}, \cite{LPR}, \cite{manin}, \cite{Art}.

This paper is devoted to integrability of system (\ref{ori}). We consider a
connection of (\ref{ori}) with a \textit{widest} class of semi-Hamiltonian
hydrodynamic-type systems%
\begin{equation}
a_{t^{k}}^{i}=\partial _{x}\frac{\partial c^{i}}{\partial a^{k}};
\label{orient}
\end{equation}%
we present a geometrical interpretation for (\ref{ori}) and linear spectral
problems (\ref{zak}) and (\ref{flow}), and describe some transformations
preserving (\ref{ori}), (\ref{zak}) and (\ref{flow}).

The celebrated WDVV equation (see, for instance, \cite{Dubr}, \cite%
{centroaffine}, \cite{Nutku}, \cite{Krich})%
\begin{equation}
\frac{\partial ^{3}F}{\partial a^{i}\partial a^{j}\partial a^{m}}\eta ^{mn}%
\frac{\partial ^{3}F}{\partial a^{n}\partial a^{k}\partial a^{s}}=\frac{%
\partial ^{3}F}{\partial a^{i}\partial a^{k}\partial a^{m}}\eta ^{mn}\frac{%
\partial ^{3}F}{\partial a^{n}\partial a^{j}\partial a^{s}},\text{ \ }j\neq k
\label{wdvv}
\end{equation}%
can be obtained from (\ref{ori}) by the potential reduction (where $\eta
^{ks}$ is a constant nondegenerate symmetric metric)%
\begin{equation}
c^{i}=\eta ^{im}\frac{\partial F}{\partial a^{m}}.  \label{reduc}
\end{equation}%
In this paper, we follow \cite{Dubr}, and step by step unravel the
relationship among (\ref{ori}) and the widest class of the so-called
conjugate curvilinear coordinate nets determined by (see, for instance, \cite%
{Darboux})%
\begin{eqnarray}
\partial _{i}\beta _{jk} &=&\beta _{ji}\beta _{ik},\text{ \ \ }i\neq j\neq k,
\label{darboux} \\
&&  \notag \\
\delta \beta _{ik} &=&0,\text{ \ \ }i\neq k,  \label{delta}
\end{eqnarray}%
where $\partial _{i}=$ $\partial /\partial r^{i},\delta =\Sigma \partial _{m}
$ is the so-called shift symmetry operator, and the rotation coefficients $%
\beta _{ik}$ depend on $N$ Riemann invariants $r^{k}$. The first subset of the
above equations is the famous Darboux system, while the second condition
means that the rotation coefficients depend only on differences of the Riemann
invariants. For the sake of simplicity we shall call these conjugate curvilinear
coordinate nets \textit{symmetry consistent}.\looseness=-1

The paper is organized as follows. In Section \textbf{2} we introduce the
metric and the basic set of solutions of linear systems (determining the
symmetry consistent conjugate curvilinear coordinate nets) whose
compatibility conditions imply the oriented associativity equations. In
Section \textbf{3} we construct the transformation from the symmetry
consistent conjugate curvilinear coordinate nets to the oriented
associativity equations. In Section \textbf{4} we construct the
relationship between linear spectral problems for the oriented associativity
equations and the symmetry consistent conjugate curvilinear coordinate nets.
In Section \textbf{5} we equip the oriented associativity equations by the
unity condition. We concentrate here on the three-component case. In Section
\textbf{6} we prove that these oriented associativity equations are a
Hamiltonian system. In Section \textbf{7} we prove that the oriented
associativity equations can be interpreted as a system of equations
describing $N$-component position vector of a hypersurface in centroaffine
geometry. In Section \textbf{8} we construct the inverse transformation
from the oriented associativity equations to the symmetry consistent
conjugate curvilinear coordinate nets. In Section \textbf{9} we construct
an infinite set of particular solutions for semi-Hamiltonian hydrodynamic
type systems whose rotation coefficients depend on differences of the Riemann
invariants only. Finally, in Section \textbf{10} we just emphasize the relations
among these very important systems.\looseness=-1

\section{Linear Spectral Problems. Basic Sets of Solutions}

Consider two linear systems%
\begin{equation}
\partial _{i}H_{k}=\beta _{ik}H_{i},\text{ \ \ \ }\partial _{i}\psi
_{k}=\beta _{ki}\psi _{i},\text{ \ \ }i\neq k,  \label{a}
\end{equation}%
whose rotation coefficients $\beta _{ik}$ depend on differences of the Riemann
invariants $r^{k}$ only (see (\ref{delta})). This means that a particular
set of solutions $H_{k},\psi _{i}$ satisfies two extra equations%
\begin{equation}
\delta H_{i}=\lambda H_{i},\text{ \ \ \ }\delta \psi _{i}=\lambda \psi _{i}.
\label{b}
\end{equation}%
The first set of compatibility conditions $\partial _{j}(\partial
_{i}H_{k})=\partial _{i}(\partial _{j}H_{k}),\partial _{j}(\partial _{i}\psi
_{k})=\partial _{i}(\partial _{j}\psi _{k})$ leads to a full set of equations
(\ref{darboux}) describing conjugate curvilinear coordinate nets, while the
second compatibility conditions $\partial _{j}(\delta H_{k})=\delta
(\partial _{j}H_{k}),\partial _{j}(\delta \psi _{k})=\delta (\partial
_{j}\psi _{k})$ yield (\ref{delta}).

\textbf{Remark}: The symmetry consistent conjugate curvilinear coordinate
nets are well known in classical differential geometry (see e.g.\
\cite{Darboux} and \cite{Tsar}). Moreover, system (\ref{darboux}), (\ref%
{delta}) was derived more recently in the context of algebro-geometric
solutions for multidimensional integrable systems (see \cite{Du77}). It is
interesting to note that this system also arises in quantum statistical
physics, see Slavnov \cite{Sla}, and in theory of the discrete analogue of
conjugate curvilinear coordinate nets known as D-invariant lattices (see
\cite{Doliwa}).

In this case $N$ infinite series of solutions of systems (\ref{a}) can be
recursively found by quadratures (see \cite{Tsar})%
\begin{equation}
\delta H_{k}^{(n+1,s)}=H_{k}^{(n,s)},\text{ \ \ }\delta \psi
_{i}^{(n+1,s)}=\psi _{i}^{(n,s)},\text{ \ }s=1,2,...,N,\text{\ \ }n=0,1,...,
\label{c}
\end{equation}%
where $\delta \psi _{i}^{(0,s)}=0$ and $\delta H_{i}^{(0,s)}=0$.

Choose $N$ particular solutions $H_{(k)i}\equiv H_{i}^{(0,k)}$ as the
\textit{basic} set of solutions. Then introduce a non-degenerate (and
non-constant in the generic case) symmetric metric%
\begin{equation}
\bar{g}_{ik}=\overset{N}{\underset{m=1}{\sum }}H_{(i)m}H_{(k)m}.
\label{metrics}
\end{equation}

\textbf{Lemma}: $N$ \textit{particular solutions} $\psi _{i}^{(n)}=\psi
_{i}^{(0,n)}$ \textit{can be chosen in the form}%
\begin{equation}
\psi _{i}^{(s)}=\bar{g}^{sn}H_{(n)i},  \label{psih}
\end{equation}%
\textit{where} $\bar{g}^{sn}$ \textit{is an inverse metric for} $\bar{g}_{ik}$.

\textbf{Proof}: Substituting (\ref{psih}) into the second equation of (\ref%
{a}) yields%
\begin{equation*}
\partial _{i}(\bar{g}^{sn}H_{(n)k})=\beta _{ki}\bar{g}^{sn}H_{(n)i},\text{ \
\ }i\neq k.
\end{equation*}%
Upon removing the parentheses on the l.h.s. and multiplying both sides by $\bar{%
g}_{js}$ the above relations boil down to%
\begin{equation*}
H_{(n)k}\bar{g}^{ns}\partial _{i}\bar{g}_{sj}=(\beta _{ik}-\beta
_{ki})H_{(j)i},\text{ \ \ }i\neq k.
\end{equation*}%
Taking into account (recall that $\delta H_{(s)i}=0$) the equations%
\begin{equation}
\partial _{i}H_{(s)i}=-\underset{m\neq i}{\sum }\beta _{mi}H_{(s)m},\text{ \
\ \ }\partial _{i}\bar{g}_{jk}=\underset{m\neq i}{\sum }(\beta _{im}-\beta
_{mi})(H_{(j)i}H_{(k)m}+H_{(k)i}H_{(j)m}),  \label{pohod}
\end{equation}%
one obtains an identity. The theorem is proved.

Then, obviously, three additional identities (where $\delta _{k}^{i}$ and $%
\delta _{ik}$ are the Kronecker symbols)%
\begin{equation}
\bar{g}^{ik}=\overset{N}{\underset{m=1}{\sum }}\psi _{m}^{(i)}\psi
_{m}^{(k)},\text{ \ \ }\delta _{k}^{i}=\overset{N}{\underset{m=1}{\sum }}%
\psi _{m}^{(i)}H_{(k)m},\text{ \ \ }\delta _{ik}=\overset{N}{\underset{m=1}{%
\sum }}\psi _{i}^{(m)}H_{(m)k}  \label{metr}
\end{equation}%
hold. In the Egorov case $\beta _{ik}=\beta _{ki}$ and the metric $\bar{g}%
^{ik}=\mathrm{const}$ (see (\ref{pohod})).

\section{Reconstruction of Oriented Associativity Equations}

In contrast with the previous section, we consider $N$ Riemann invariants $%
r^{k}$ as functions of $N$ independent variables $t^{n}$, i.e. we introduce $%
N$ commuting hydrodynamic-type systems%
\begin{equation}
r_{t^{k}}^{i}=\frac{H_{(k)i}}{\bar{H}_{i}}r_{x}^{i},  \label{rim}
\end{equation}%
where $\bar{H}_{i}$ is an arbitrary solution of the first linear system in (%
\ref{a}). These hydrodynamic-type systems are semi-Hamiltonian (i.e. possess
infinite set of conservation laws parameterized by $N$ arbitrary functions
of a single variable, see \cite{Tsar}). In such a case, they can be written
in the conservative form%
\begin{equation}
\partial _{t^{k}}h=\partial _{x}g_{k},  \label{sohr}
\end{equation}%
where $\partial _{i}h=\psi _{i}\bar{H}_{i},\partial _{i}g_{k}=\psi
_{i}H_{(k)i}$ and $\psi _{i}$ is an arbitrary solution (parameterized by $N$
arbitrary functions of a single variable, see \cite{Tsar}) of the second
system in (\ref{a}). Introduce $N$ conservation law densities $a^{k}$ such
that $\partial _{i}a^{k}=\psi _{i}^{(k)}\bar{H}_{i}$ and $N$ conservation
law densities $c^{k}$ such that $\partial _{i}c^{k}=\bar{\psi}_{i}^{(k)}\bar{%
H}_{i}$, where $\delta \bar{\psi}_{i}^{(k)}=\psi _{i}^{(k)}$ (i.e. $\bar{\psi%
}_{i}^{(k)}\equiv \psi _{i}^{(1,k)}$, see (\ref{c}) and (\ref{psih})).

\textbf{Theorem}: $N$ \textit{commuting hydrodynamic-type systems} (\ref{rim}%
) \textit{can be written in the conservative form} (\ref{orient}), \textit{%
whose compatibility conditions are oriented associativity equations} (\ref%
{ori}).

\textbf{Proof}: The relation $\bar{\psi}_{i}^{(k)}=c_{s}^{k}\psi
_{i}^{(s)}$ follows from%
\begin{equation*}
dc^{k}=\overset{N}{\underset{m=1}{\sum }}\bar{\psi}_{m}^{(k)}\bar{H}%
_{m}dr^{m}=c_{s}^{k}da^{s}=c_{s}^{k}\overset{N}{\underset{m=1}{\sum }}\psi
_{m}^{(s)}\bar{H}_{m}dr^{m},
\end{equation*}%
where $c_{k}^{i}\equiv \partial c^{i}/\partial a^{k}$ (see (\ref{orient})).
Taking into account the second identity from (\ref{metr}), one can obtain%
\begin{equation}
c_{k}^{i}=\overset{N}{\underset{m=1}{\sum }}\bar{\psi}_{m}^{(i)}H_{(k)m}.
\label{cik}
\end{equation}%
Then (see (\ref{a}))%
\begin{eqnarray*}
\partial _{i}c_{j}^{k} &=&\partial _{i}\left( \bar{\psi}_{i}^{(k)}H_{(j)i}+%
\underset{m\neq i}{\sum }\bar{\psi}_{m}^{(k)}H_{(j)m}\right) \\
&& \\
&=&\bar{\psi}_{i}^{(k)}\partial _{i}H_{(j)i}+H_{(j)i}\partial _{i}\bar{\psi}%
_{i}^{(k)}+H_{(j)i}\underset{m\neq i}{\sum }\beta _{im}\bar{\psi}_{m}^{(k)}+%
\bar{\psi}_{i}^{(k)}\underset{m\neq i}{\sum }\beta _{mi}H_{(j)m} \\
&& \\
&=&\bar{\psi}_{i}^{(k)}\left( \partial _{i}H_{(j)i}+\underset{m\neq i}{\sum }%
\beta _{mi}H_{(j)m}\right) +H_{(j)i}\left( \partial _{i}\bar{\psi}_{i}^{(k)}+%
\underset{m\neq i}{\sum }\beta _{im}\bar{\psi}_{m}^{(k)}\right) .
\end{eqnarray*}%
If we take into account that the expression in the first brackets vanishes
because $\delta H_{(j)i}=0$, and the expression in the second brackets is
nothing but $\psi _{i}^{(k)}$ (since $\delta \bar{\psi}_{i}^{(k)}=\psi
_{i}^{(k)}$), then one can conclude that%
\begin{equation}
\partial _{i}c_{j}^{k}=\psi _{i}^{(k)}H_{(j)i}.  \label{deriv}
\end{equation}%
On the other hand, if the hydrodynamic-type systems (\ref{rim}) possess $N$
conservation laws (\ref{orient}), then%
\begin{equation*}
\partial _{m}a^{i}\cdot r_{t^{k}}^{m}=\partial _{m}c_{k}^{i}\cdot r_{x}^{m}.
\end{equation*}%
Substituting the r.h.s. of (\ref{rim}) for $r_{t^{k}}^{i}$ leads to
(recall that $\partial _{i}a^{k}=\psi _{i}^{(k)}\bar{H}_{i}$)%
\begin{equation*}
\partial _{i}c_{j}^{k}=\frac{H_{(j)i}}{\bar{H}_{i}}\psi _{i}^{(k)}\bar{H}%
_{i}=\psi _{i}^{(k)}H_{(j)i},
\end{equation*}%
which coincide with (\ref{deriv}). The theorem is proved.

\textbf{Remark}: The second derivatives of the functions $c^{i}$ with respect to
the field variables $a^{j},a^{k}$ can be easily derived from (\ref{deriv}).
Indeed, the relations%
\begin{equation*}
\partial _{i}c_{j}^{k}=c_{js}^{k}\partial _{i}a^{s}=c_{js}^{k}\psi _{i}^{(s)}%
\bar{H}_{i}=\psi _{i}^{(k)}H_{(j)i}
\end{equation*}%
lead (upon multiplying the third and fourth blocks of the above expression by
the ratio $H_{(p)i}/\bar{H}_{i}$, and summing according to the second
formula from (\ref{metr})) to
\begin{equation}
c_{jk}^{i}=\overset{N}{\underset{m=1}{\sum }}\frac{\psi
_{m}^{(i)}H_{(j)m}H_{(k)m}}{\bar{H}_{m}},  \label{odindva}
\end{equation}%
which, obviously, satisfy (\ref{ori}) by virtue of (\ref{metr}). Its
symmetric form%
\begin{equation*}
c_{ijk}=\bar{g}_{is}c_{jk}^{s}=\overset{N}{\underset{m=1}{\sum }}\frac{%
H_{(i)m}H_{(j)m}H_{(k)m}}{\bar{H}_{m}}
\end{equation*}%
is well known in the theory of WDVV associativity equations and Frobenius
manifolds, where such expressions for $c_{ijk}$ can be found using the theory
of meromorphic differentials on algebraic Riemann surfaces (see, for
instance, \cite{Dubr} and \cite{Krich}).

\textbf{Remark}: Since%
\begin{equation*}
\overset{N}{\underset{m=1}{\sum }}\partial _{m}a^{k}\cdot \frac{\partial
r^{m}}{\partial a^{s}}=\delta _{s}^{k},\text{ \ \ }\overset{N}{\underset{m=1}%
{\sum }}\partial _{k}a^{m}\cdot \frac{\partial r^{s}}{\partial a^{m}}=\delta
_{k}^{s},
\end{equation*}%
one can easily derive (recall (\ref{metr}) and $\partial _{i}a^{k}=\psi
_{i}^{(k)}\bar{H}_{i}$)%
\begin{equation}
\frac{\partial r^{i}}{\partial a^{k}}=\frac{H_{(k)i}}{\bar{H}_{i}}.
\label{rimder}
\end{equation}%
Thus, the characteristic velocities $v_{(k)}^{i}(\mathbf{a})\equiv H_{(k)i}/%
\bar{H}_{i}$ (see (\ref{rim})) of $N$ commuting hydrodynamic-type systems (%
\ref{orient}) are nothing but $\partial r^{i}/\partial a^{k}$. Thus, the
compatibility conditions $\partial v_{(k)}^{i}/\partial a^{j}=\partial
v_{(j)}^{i}/\partial a^{k}$ should be satisfied. Indeed, multiplying both
sides by $a_{i}^{p}$ and summing over $i$, one obtains
an identity, which follows from%
\begin{equation*}
\overset{N}{\underset{m=1}{\sum }}v_{(k)}^{m}\frac{\partial a_{m}^{p}}{%
\partial a^{j}}=\overset{N}{\underset{m=1}{\sum }}v_{(j)}^{m}\frac{\partial
a_{m}^{p}}{\partial a^{k}},
\end{equation*}%
where $\partial a_{m}^{i}/\partial a^{k}=a_{ms}^{i}v_{(k)}^{s}$. Thus, we
conclude that%
\begin{equation}
\frac{\partial r^{i}}{\partial a^{k}}=v_{(k)}^{i}(\mathbf{a}).  \label{karak}
\end{equation}

Any semi-Hamiltonian hydrodynamic-type system (see \cite{Tsar})%
\begin{equation}
a_{t}^{i}=v_{k}^{i}a_{x}^{k},  \label{hds}
\end{equation}%
is associated with the non-degenerate metric tensor $\bar{g}^{ik}$. The
necessary and sufficient conditions for existence of this tensor are given
by (here $\nabla _{k}$ is the covariant derivative) the Tsarev lemma (see \cite%
{Tsar})%
\begin{equation}
\bar{g}^{ik}v_{k}^{j}=\bar{g}^{jk}v_{k}^{i},\text{\ \ \ }\nabla
_{i}v_{j}^{k}=\nabla _{j}v_{i}^{k}.  \label{ham}
\end{equation}%
However, for hydrodynamic-type systems (\ref{orient}) we should not solve
this system, because we already know that the metric tensor $\bar{g}_{ij}$
in the field variables $a^{k}$ is given by (\ref{metrics}).
Indeed, we have (see (\ref{metrics}), (\ref{metr}))
\begin{equation*}
ds^{2}=\overset{N}{\underset{i=1}{\sum }}\overset{N}{\underset{k=1}{\sum }}%
\bar{g}_{ik}da^{i}da^{k}=\overset{N}{\underset{i=1}{\sum }}\overset{N}{%
\underset{k=1}{\sum }}\overset{N}{\underset{s=1}{\sum }}\overset{N}{\underset%
{n=1}{\sum }}\left( \overset{N}{\underset{m=1}{\sum }}H_{(i)m}H_{(k)m}%
\right) (\psi _{s}^{(i)}\bar{H}_{s}dr^{s})(\psi _{n}^{(k)}\bar{H}_{n}dr^{n})
\end{equation*}%
\begin{equation*}
=\overset{N}{\underset{s=1}{\sum }}\overset{N}{\underset{n=1}{\sum }}\overset%
{N}{\underset{m=1}{\sum }}\left( \overset{N}{\underset{i=1}{\sum }}\psi
_{s}^{(i)}H_{(i)m}\right) \left( \overset{N}{\underset{k=1}{\sum }}\psi
_{n}^{(k)}H_{(k)m}\right) \bar{H}_{n}\bar{H}_{s}dr^{s}dr^{n}=\overset{N}{%
\underset{m=1}{\sum }}\bar{H}_{m}^{2}(dr^{m})^{2}.
\end{equation*}%
So, we arrive at the conclusion that the diagonal metric $g_{kk}=\bar{H}%
_{k}^{2}$ in the Riemann invariants, in perfect agreement with Tsarev's
definition of semi-Hamiltonian metric (see \cite{Tsar}).

Anyway, taking into account the expression for the covariant derivative $%
\nabla _{l}v_{k}^{i}=\partial _{l}v_{k}^{i}-\Gamma _{lk}^{m}v_{m}^{i}+\Gamma
_{lm}^{i}v_{k}^{m}$, conditions (\ref{ham}) for hydrodynamic-type systems (%
\ref{orient}) reduce to a more compact form%
\begin{equation}
\bar{g}^{ik}c_{ks}^{j}=\bar{g}^{jk}c_{ks}^{i},\text{\ \ \ }\Gamma
_{im}^{k}c_{js}^{m}=\Gamma _{jm}^{k}c_{is}^{m}.  \label{dvegrupy}
\end{equation}%
By virtue of (\ref{metr}) and (\ref{odindva}), the first group of equations
yields an identity. To consider the second group of equations, at first we
should compute%
\begin{equation*}
\Gamma _{jk}^{i}=\frac{1}{2}\bar{g}^{im}\left( \frac{\partial \bar{g}_{mk}}{%
\partial a^{j}}+\frac{\partial \bar{g}_{mj}}{\partial a^{k}}-\frac{\partial
\bar{g}_{jk}}{\partial a^{m}}\right) .
\end{equation*}%
Taking into account (\ref{pohod}), (\ref{metr}) and (\ref{rimder}) yields%
\begin{equation*}
\Gamma _{jk}^{i}=\frac{1}{2}\overset{N}{\underset{m=1}{\sum }}\overset{N}{%
\underset{s=1}{\sum }}\bar{g}^{im}\left( \partial _{s}\bar{g}_{mk}\frac{%
\partial r^{s}}{\partial a^{j}}+\partial _{s}\bar{g}_{mj}\frac{\partial r^{s}%
}{\partial a^{k}}-\partial _{s}\bar{g}_{jk}\frac{\partial r^{s}}{\partial
a^{m}}\right) =\overset{N}{\underset{s=1}{\sum }}\frac{H_{(j)s}H_{(k)s}}{%
\bar{H}_{s}}\overset{N}{\underset{p=1}{\sum }}(\beta _{sp}-\beta _{ps})\psi
_{p}^{(i)}.
\end{equation*}%
Then, upon taking into account (\ref{metr}) and (\ref{odindva}) once again, the second
group of equations in (\ref{dvegrupy}) becomes an identity.\looseness=-1

\section{Linear Spectral Problems for Oriented Associativity Equations}

As it was already mentioned in Introduction, the oriented associativity equations (\ref%
{ori}) can be obtained as compatibility conditions of the scalar linear
spectral problem (\ref{zak})%
\begin{equation}
h_{ik}=\lambda c_{ik}^{s}h_{s},  \label{zaks}
\end{equation}%
where we denote $c_{jk}^{i}\equiv \partial ^{2}c^{i}/\partial a^{j}\partial
a^{k},h_{i}\equiv \partial h/\partial a^{i},h_{ik}\equiv \partial
^{2}h/\partial a^{i}\partial a^{k}$.

\textbf{Lemma}: \textit{The function} $h(\mathbf{a})$ \textit{is a
generating function of conservation law densities} (\textit{see} (\ref{sohr}%
)).

\textbf{Proof}: If $h(\mathbf{a})$ is a conservation law density, then (see (%
\ref{rimder}) and cf. (\ref{cik}))%
\begin{equation*}
h_{k}=\overset{N}{\underset{m=1}{\sum }}\partial _{m}h\cdot \frac{\partial
r^{m}}{\partial a^{k}}=\overset{N}{\underset{m=1}{\sum }}\psi _{m}\bar{H}%
_{m}\cdot \frac{H_{(k)m}}{\bar{H}_{m}}=\overset{N}{\underset{m=1}{\sum }}%
\psi _{m}H_{(k)m}.
\end{equation*}%
Also, taking into account (\ref{zaks}) and (\ref{deriv}), one can compute
\begin{equation*}
dh_{k}=h_{km}da^{m}=\lambda c_{km}^{s}h_{s}da^{m}=\lambda h_{s}dc_{k}^{s}
\end{equation*}%
\begin{equation*}
=\lambda \overset{N}{\underset{s=1}{\sum }}\overset{N}{\underset{n=1}{\sum }}%
\overset{N}{\underset{m=1}{\sum }}(\psi _{n}H_{(s)n})(\psi
_{m}^{(s)}H_{(k)m}dr^{m})=\lambda \overset{N}{\underset{n=1}{\sum }}\overset{%
N}{\underset{m=1}{\sum }}\left( \overset{N}{\underset{s=1}{\sum }}%
H_{(s)n}\psi _{m}^{(s)}\right) \psi _{n}H_{(k)m}dr^{m},
\end{equation*}%
i.e. we have
\begin{equation*}
dh_{k}=\lambda \overset{N}{\underset{m=1}{\sum }}\psi _{m}H_{(k)m}dr^{m}.
\end{equation*}%
So, we arrive at the relation%
\begin{equation*}
\partial _{i}\left( \overset{N}{\underset{m=1}{\sum }}\psi
_{m}H_{(k)m}\right) =\lambda \psi _{i}H_{(k)i},
\end{equation*}%
which immediately reduces to the second equation in (\ref{b}). The lemma is
proved.

Now, consider the commuting hydrodynamic-type system ($\tau $ is a group
parameter)%
\begin{equation}
r_{\tau }^{i}=\frac{H_{i}}{\bar{H}_{i}}r_{x}^{i},\text{\ }\Leftrightarrow
\text{ \ }a_{\tau }^{k}=b_{x}^{k},  \label{flux}
\end{equation}%
where $H_{i}$ is an arbitrary solution of first linear system (\ref{a}) and $%
\partial _{i}b^{k}=\psi _{i}^{(k)}H_{i}$. As it was mentioned in
Introduction, oriented associativity equations (\ref{ori}) can be obtained
as compatibility conditions of the vector linear spectral problem (\ref{flow}%
)%
\begin{equation}
b_{k}^{i}=\lambda c_{ks}^{i}b^{s},  \label{bik}
\end{equation}%
where we denote $b_{k}^{i}\equiv \partial b^{i}/\partial a^{k}$.

\textbf{Lemma}: \textit{The functions}
\begin{equation*}
b^{k}(\mathbf{a})=\lambda ^{-1}\overset{N}{\underset{m=1}{\sum }}\psi
_{m}^{(k)}H_{m}
\end{equation*}%
\textit{are conservation law fluxes for generating functions of commuting
flows} (\ref{flux}).

\textbf{Proof}: Taking into account (\ref{bik}), (\ref{deriv}) and (\ref%
{metr}), we find that
\begin{equation*}
db^{k}=\lambda c_{ns}^{k}b^{s}da^{n}=\lambda b^{s}dc_{s}^{k}=\overset{N}{%
\underset{s=1}{\sum }}\left( \overset{N}{\underset{m=1}{\sum }}\psi
_{m}^{(s)}H_{m}\right) \left( \overset{N}{\underset{n=1}{\sum }}\psi
_{n}^{(k)}H_{(s)n}dr^{n}\right)
\end{equation*}%
\begin{equation*}
=\overset{N}{\underset{m=1}{\sum }}\overset{N}{\underset{n=1}{\sum }}\left(
\overset{N}{\underset{s=1}{\sum }}\psi _{m}^{(s)}H_{(s)n}\right) \psi
_{n}^{(k)}H_{m}dr^{n}=\overset{N}{\underset{m=1}{\sum }}\psi
_{m}^{(k)}H_{m}dr^{m}.
\end{equation*}%
Thus, indeed, $\partial _{i}b^{k}=\psi _{i}^{(k)}H_{i}$. On the other hand,
we have%
\begin{equation*}
\lambda \partial _{i}b^{k}=\underset{m\neq i}{\sum }\psi _{m}^{(k)}\partial
_{i}H_{m}+\psi _{i}^{(k)}\partial _{i}H_{i}+\underset{m\neq i}{\sum }%
H_{m}\partial _{i}\psi _{m}^{(k)}+H_{i}\partial _{i}\psi _{i}^{(k)},
\end{equation*}%
i.e.%
\begin{equation*}
\lambda \partial _{i}b^{k}=H_{i}\underset{m\neq i}{\sum }\beta _{im}\psi
_{m}^{(k)}+\psi _{i}^{(k)}\left( \delta H_{i}-\underset{m\neq i}{\sum }\beta
_{mi}H_{m}\right) +\psi _{i}^{(k)}\underset{m\neq i}{\sum }\beta
_{mi}H_{m}+H_{i}\left( \delta \psi _{i}^{(k)}-\underset{m\neq i}{\sum }\beta
_{im}\psi _{m}^{(k)}\right) .
\end{equation*}%
Since the first sum equals to the fourth sum, and the second sum equals to
the third sum, we arrive at the conclusion that $\lambda \partial
_{i}b^{k}=\psi _{i}^{(k)}\delta H_{i}$ (recall that $\delta \psi
_{i}^{(k)}=0 $), which agrees with the previously computed $\partial
_{i}b^{k}=\psi _{i}^{(k)}H_{i}$, if and only if $H_{i}$ satisfies the first
equation in the linear spectral problem (\ref{b}). The lemma is proved.\looseness=-1

\textbf{Remark}: One commuting flow (\ref{flux}) can be found without
integration, i.e. ($y$ is a group parameter)%
\begin{equation}
a_{y}^{i}=(a^{s}c_{s}^{i}-c^{i})_{x}.  \label{potok}
\end{equation}%
Indeed (see (\ref{odindva})), we have
\begin{equation*}
\partial _{i}(a^{s}c_{s}^{k}-c^{k})=a^{s}c_{sn}^{k}\partial
_{i}a^{n}=a^{s}c_{sn}^{k}\psi _{i}^{(n)}\bar{H}_{i}=a^{s}\overset{N}{%
\underset{n=1}{\sum }}\overset{N}{\underset{m=1}{\sum }}\frac{\psi
_{m}^{(k)}H_{(s)m}H_{(n)m}}{\bar{H}_{m}}\psi _{i}^{(n)}\bar{H}_{i}
\end{equation*}%
\begin{equation*}
=a^{s}\overset{N}{\underset{m=1}{\sum }}\frac{\psi _{m}^{(k)}H_{(s)m}}{\bar{H%
}_{m}}\left( \overset{N}{\underset{n=1}{\sum }}\psi
_{i}^{(n)}H_{(n)m}\right) \bar{H}_{i}=a^{s}\psi _{i}^{(k)}H_{(s)i}.
\end{equation*}%
Since $\partial _{i}(a^{s}c_{s}^{k}-c^{k})=\psi _{i}^{(k)}\tilde{H}_{i}$,
where $\tilde{H}_{i}$ is some solution of the first system in (\ref{a}), we
conclude that $\tilde{H}_{i}=a^{s}H_{(s)i}$. Substituting this expression into
the first system in (\ref{a}) leads to an identity, and the substitution
into the first equation in (\ref{b}) yields%
\begin{equation*}
\delta \tilde{H}_{i}=H_{(s)i}\delta a^{s}=\overset{N}{\underset{s=1}{\sum }}%
H_{(s)i}\overset{N}{\underset{m=1}{\sum }}\psi _{m}^{(s)}\bar{H}_{m}=\overset%
{N}{\underset{m=1}{\sum }}\left( \overset{N}{\underset{s=1}{\sum }}\psi
_{m}^{(s)}H_{(s)i}\right) \bar{H}_{m}=\bar{H}_{i}.
\end{equation*}%
Thus, the family of hydrodynamic-type systems (\ref{orient}) has a simple
commuting flow (\ref{potok}), which in the diagonal form (see the first
equation in (\ref{flux})) has characteristic velocities $\tilde{H}_{i}/\bar{H%
}_{i}$ such that $\delta \tilde{H}_{i}=\bar{H}_{i}$.

\section{Reduction to Canonical Form}

Hydrodynamic-type systems (\ref{orient}) have  $N$ additional natural conservation
laws%
\begin{equation}
c_{t^{k}}^{i}=\partial _{x}\frac{\partial Q^{i}}{\partial a^{k}}.  \label{si}
\end{equation}%
Indeed, suppose that the hydrodynamic-type systems (\ref{orient}) possess $N$
additional conservation laws $c_{t^{k}}^{i}=\partial _{x}Q_{k}^{i}$. Then $%
dQ_{k}^{i}=c_{m}^{i}dc_{k}^{m}$, i.e. $Q_{ks}^{i}=c_{m}^{i}c_{ks}^{m}$.
Thus, we conclude that $Q_{ks}^{i}=Q_{sk}^{i}$ and $N$ functions $Q^{i}$
determine the fluxes of these conservation laws. The compatibility conditions
$\partial Q_{ks}^{i}/\partial a^{j}=\partial Q_{js}^{i}/\partial a^{k}$ hold by virtue of (\ref{ori}).

Rewrite $N$ commuting hydrodynamic-type systems (\ref{orient}) in the
differential form, i.e. (see (\ref{cik})),
\begin{equation*}
dz^{i}=a^{i}dx+\overset{N}{\underset{k=1}{\sum }}c_{k}^{i}(\mathbf{a}%
)dt^{k}\equiv a^{i}dx+\overset{N}{\underset{m=1}{\sum }}\overset{N}{\underset%
{k=1}{\sum }}\bar{\psi}_{m}^{(i)}H_{(k)m}dt^{k}.
\end{equation*}%
Thus, we see that functions $c_{k}^{i}(\mathbf{a})$ do not depend on the
choice of Lam\`{e} coefficients $\bar{H}_{k}$, while (recall again) $%
\partial _{i}a^{k}=\psi _{i}^{(k)}\bar{H}_{i}$. Below we shall consider the
reduced version, i.e.
\begin{equation*}
dz^{i}=\overset{N}{\underset{m=1}{\sum }}\overset{N}{\underset{k=1}{\sum }}%
\bar{\psi}_{m}^{(i)}H_{(k)m}dt^{k}.
\end{equation*}%
In such a case, we pick the first ``time"
variable $t^{1}$ and choose the new field variables $\tilde{a}^{i}\equiv
c_{1}^{i}(\mathbf{a})$, and then instead of (\ref{orient}) we shall consider just $%
N-1$ commuting flows%
\begin{equation}
\tilde{a}_{t^{k}}^{i}=\partial _{t^{1}}\frac{\partial \tilde{c}^{i}}{%
\partial \tilde{a}^{k}},  \label{reduk}
\end{equation}%
where $\partial \tilde{c}^{i}/\partial \tilde{a}^{k}=\partial c^{i}/\partial
a^{k},k=2,3,...,N$. Thus, all functions $\tilde{c}^{i}$ can be found in
quadratures, i.e.%
\begin{equation*}
d\tilde{c}^{i}=\overset{N}{\underset{m=1}{\sum }}c_{m}^{i}dc_{1}^{m}.
\end{equation*}%
Indeed, rewrite the conservation laws (\ref{si}) in the differential form%
\begin{equation*}
dy^{i}=c^{i}(\mathbf{a})dx+\overset{N}{\underset{k=1}{\sum }}Q_{k}^{i}(%
\mathbf{a})dt^{k}=c^{i}(\mathbf{a})dx+Q_{1}^{i}(\mathbf{a})dt^{1}+\overset{N}%
{\underset{k=2}{\sum }}Q_{k}^{i}(\mathbf{a})dt^{k}.
\end{equation*}%
Choose $\tilde{c}^{i}=Q_{1}^{i}(\mathbf{a})$, then $d\tilde{c}%
^{i}=Q_{1s}^{i}da^{s}=c_{m}^{i}c_{1s}^{m}da^{s}=c_{m}^{i}dc_{1}^{m}=c_{m}^{i}d%
\tilde{a}^{m}$. So, $\partial \tilde{c}^{i}/\partial \tilde{a}^{k}=\partial
c^{i}/\partial a^{k}$.

Obviously, the hydrodynamic-type systems (\ref{reduk}) can be written in the
diagonal form%
\begin{equation}
r_{t^{k}}^{i}=\frac{H_{(k)i}}{H_{(1)i}}r_{t^{1}}^{i}.  \label{short}
\end{equation}%
Indeed, the first of these commuting flows (\ref{rim}) can be written in the
form%
\begin{equation*}
r_{x}^{i}=\frac{\bar{H}_{i}}{H_{(1)i}}r_{t^{1}}^{i}.
\end{equation*}%
Then all other commuting flows (\ref{rim}) reduce to (\ref{short}) upon substituting for $r_{x}^{i}$ from the above formula.

Below we shall omit tildes over field variables $a^{k}$ and $c^{n}$, because
we are going to consider just the oriented associativity equations
supplemented with the so-called \textquotedblleft unity\textquotedblright\ condition (cf.
\cite{Dubr})
\begin{equation}
c_{1k}^{i}=\delta _{k}^{i},  \label{unity}
\end{equation}%
which is equivalent to considering a canonical set of $N-1$ commuting flows (%
\ref{reduk}). We shall call such oriented associativity equations \textit{%
normalized}.

Thus, the conservative representations (\ref{reduk}) reduce to the form%
\begin{equation*}
a_{t^{k}}^{1}=\partial _{t^{1}}u_{k}^{1},\text{\ \ \ }a_{t^{k}}^{i}=\partial
_{t^{1}}(a^{1}\delta _{k}^{i}+u_{k}^{i}),\text{ \ \ }i,k=2,3,...,N,
\end{equation*}%
where the new unknown functions $u^{n}(a^{2},a^{3},...,a^{N})$ appear from
integration of (\ref{unity}), i.e.%
\begin{equation}
c^{1}=\frac{1}{2}(a^{1})^{2}+u^{1},\text{ \ \ }c^{k}=a^{1}a^{k}+u^{k},\text{
\ }k=2,3,...,N.  \label{canon}
\end{equation}%
In this case, the shift symmetry operator $\delta $ can be easily expressed
via the field variables $a^{k}$ instead of the Riemann invariants $r^{n}$,
i.e.%
\begin{equation}
\delta =\overset{N}{\underset{m=1}{\sum }}\frac{\partial }{\partial r^{m}}=%
\overset{N}{\underset{k=1}{\sum }}\left( \overset{N}{\underset{m=1}{\sum }}%
\frac{\partial a^{k}}{\partial r^{m}}\right) \frac{\partial }{\partial a^{k}}%
=\overset{N}{\underset{k=1}{\sum }}\left( \overset{N}{\underset{m=1}{\sum }}%
\psi _{m}^{(k)}H_{(1)m}\right) \frac{\partial }{\partial a^{k}}=\frac{%
\partial }{\partial a^{1}}.  \label{dda}
\end{equation}%
Thus, two equations of scalar and vector linear spectral problems (\ref{zaks}%
) and (\ref{bik}), namely%
\begin{equation}
h_{1}=\lambda h,\text{ \ \ }b_{1}^{k}=\lambda b^{k}  \label{shift}
\end{equation}%
coincide with the eigenvalue problem for the above shift symmetry operator,
while the remaining equations in (\ref{zaks}) and (\ref{bik}) become,
respectively,
\begin{equation}
h_{ik}=\lambda \overset{N}{\underset{s=1}{\sum }}u_{ik}^{s}h_{s},\text{\ }%
i,k=2,3,\dots,N,  \label{lax}
\end{equation}%
\begin{equation*}
b_{k}^{i}=\lambda \left( (\delta _{k}^{i}-\delta _{1}^{i}\delta
_{k}^{1})b^{1}+\overset{N}{\underset{m=2}{\sum }}u_{km}^{i}b^{m}\right) ,%
\text{ \ }i=1,2,...,N,\text{ \ }k=2,3,...,N.
\end{equation*}

\textbf{Example}: Consider the simplest nontrivial case $N=3$. Then two
commuting flows are given by (here $x=t^{1},t=t^{2},y=t^{3}$)%
\begin{eqnarray}
a_{t} &=&\partial _{x}u_{b},\text{ \ \ }b_{t}=\partial _{x}(a+v_{b}),\text{
\ \ }c_{t}=\partial _{x}w_{b};  \notag \\
&&  \label{syst} \\
a_{y} &=&\partial _{x}u_{c},\text{ \ \ }b_{y}=\partial _{x}v_{c},\text{ \ \ }%
c_{y}=\partial _{x}(a+w_{c}),  \notag
\end{eqnarray}%
where $a=a^{1},b=a^{2},c=a^{3},u=u^{1}(b,c),v=u^{2}(b,c),w=u^{3}(b,c)$ and
the subscripts indicate the corresponding partial derivatives. The
compatibility conditions for (\ref{syst}), which read $
(a_{y})_{t}=(a_{t})_{y},(b_{y})_{t}=(b_{t})_{y},(c_{y})_{t}=(c_{t})_{y}$,
lead to three algebraic equations relating the second-order derivatives%
\begin{equation*}
u_{bb}=v_{bc}w_{bb}-v_{bb}w_{bc}+w_{bc}^{2}-w_{bb}w_{cc},\text{ \ }%
u_{bc}=v_{cc}w_{bb}-v_{bc}w_{bc},\text{\ \ }%
u_{cc}=v_{bc}^{2}-v_{bb}v_{cc}+v_{cc}w_{bc}-v_{bc}w_{cc},
\end{equation*}%
which are integrable by the inverse spectral transform. The scalar Lax pair
for the latter (see (\ref{lax})) is given by%
\begin{equation*}
\left(
\begin{array}{c}
h_{a} \\
h_{b} \\
h_{c}%
\end{array}%
\right) _{b}=\lambda \left(
\begin{array}{ccc}
0 & 1 & 0 \\
u_{bb} & v_{bb} & w_{bb} \\
u_{bc} & v_{bc} & w_{bc}%
\end{array}%
\right) \left(
\begin{array}{c}
h_{a} \\
h_{b} \\
h_{c}%
\end{array}%
\right) ,\text{ \ }\left(
\begin{array}{c}
h_{a} \\
h_{b} \\
h_{c}%
\end{array}%
\right) _{c}=\lambda \left(
\begin{array}{ccc}
0 & 0 & 1 \\
u_{bc} & v_{bc} & w_{bc} \\
u_{cc} & v_{cc} & w_{cc}%
\end{array}%
\right) \left(
\begin{array}{c}
h_{a} \\
h_{b} \\
h_{c}%
\end{array}%
\right) ,
\end{equation*}%
while the adjoint Lax pair has transposed matrices, i.e.%
\begin{equation*}
\left(
\begin{array}{c}
b^{1} \\
b^{2} \\
b^{3}%
\end{array}%
\right) _{b}=\lambda \left(
\begin{array}{ccc}
0 & u_{bb} & u_{bc} \\
1 & v_{bb} & v_{bc} \\
0 & w_{bb} & w_{bc}%
\end{array}%
\right) \left(
\begin{array}{c}
b^{1} \\
b^{2} \\
b^{3}%
\end{array}%
\right) ,\text{ \ }\left(
\begin{array}{c}
b^{1} \\
b^{2} \\
b^{3}%
\end{array}%
\right) _{c}=\lambda \left(
\begin{array}{ccc}
0 & u_{bc} & u_{cc} \\
0 & v_{bc} & v_{cc} \\
1 & w_{bc} & w_{cc}%
\end{array}%
\right) \left(
\begin{array}{c}
b^{1} \\
b^{2} \\
b^{3}%
\end{array}%
\right) .
\end{equation*}

\textbf{Remark}: Both commuting hydrodynamic-type systems (\ref{lax}) also
have the same Lax matrices, i.e.%
\begin{equation*}
\left(
\begin{array}{c}
a \\
b \\
c%
\end{array}%
\right) _{t}=\left(
\begin{array}{ccc}
0 & u_{bb} & u_{bc} \\
1 & v_{bb} & v_{bc} \\
0 & w_{bb} & w_{bc}%
\end{array}%
\right) \left(
\begin{array}{c}
a \\
b \\
c%
\end{array}%
\right) _{x},\text{ \ }\left(
\begin{array}{c}
a \\
b \\
c%
\end{array}%
\right) _{y}=\left(
\begin{array}{ccc}
0 & u_{bc} & u_{cc} \\
0 & v_{bc} & v_{cc} \\
1 & w_{bc} & w_{cc}%
\end{array}%
\right) \left(
\begin{array}{c}
a \\
b \\
c%
\end{array}%
\right) _{x}.
\end{equation*}

\section{Hamiltonian Structure of Oriented Associativity Equations}

The system of quadratic equations%
\begin{equation*}
u_{bb}=v_{bc}w_{bb}-v_{bb}w_{bc}+w_{bc}^{2}-w_{bb}w_{cc},\text{ \ }%
u_{bc}=v_{cc}w_{bb}-v_{bc}w_{bc},\text{\ \ }%
u_{cc}=v_{bc}^{2}-v_{bb}v_{cc}+v_{cc}w_{bc}-v_{bc}w_{cc},
\end{equation*}%
is nothing but the oriented associativity equations in three-dimensional
case with the unity condition (\ref{unity}) (see also (\ref{canon})).
Introduce a new set of field variables $%
q^{1}=u_{bb},q^{2}=u_{bc},q^{3}=v_{bb},q^{4}=v_{bc}$, $%
q^{5}=w_{bb},q^{6}=w_{bc}$. Then this quadratic system becomes a
six-component hydrodynamic-type system%
\begin{eqnarray}
q_{c}^{1} &=&q_{b}^{2},\text{ \ \ \ \ \ \ \ \ }q_{c}^{2}=\partial _{b}\frac{%
q^{2}q^{6}+q^{1}q^{4}-q^{2}q^{3}}{q^{5}},  \notag \\
&&  \notag \\
q_{c}^{3} &=&q_{b}^{4},\text{ \ \ \ \ \ \ \ \ }q_{c}^{4}=\partial _{b}\frac{%
q^{2}+q^{4}q^{6}}{q^{5}},  \label{six} \\
&&  \notag \\
q_{c}^{5} &=&q_{b}^{6},\text{ \ \ \ \ \ \ \ \ }q_{c}^{6}=\partial _{b}\frac{%
(q^{6})^{2}-q^{3}q^{6}+q^{4}q^{5}-q^{1}}{q^{5}}.  \notag
\end{eqnarray}%
As it was already mentioned in Section \textbf{5}, in the present
paper we restrict ourselves to the generic case when all characteristic velocities $v_{(k)}^{i}(%
\mathbf{a})$ are pairwise distinct. We seek the Riemann invariants $r(a,b,c)$
for both commuting hydrodynamic-type systems (\ref{syst}) written in the
diagonal form%
\begin{equation}
r_{t}=v_{(2)}r_{x},\text{ \ \ }r_{y}=v_{(3)}r_{x}.  \label{riman}
\end{equation}%
Taking into account that $r_{a}=1$ (thanks to the shift symmetry operator $%
\delta =\partial _{a}$, see (\ref{dda})), one can obtain%
\begin{equation}
r_{b}=v_{(2)},\text{ \ \ }r_{c}=v_{(3)},  \label{roots}
\end{equation}%
where the characteristic velocities $v_{(2)}$ and $v_{(3)}$ of (\ref{syst}) are
related polynomially,
\begin{equation*}
v_{(3)}=\frac{(v_{(2)})^{2}-q^{3}v_{(2)}-q^{1}}{q^{5}},
\end{equation*}%
while $v_{(2)}$ satisfies the characteristic equation (\ref{korni}) for the
first hydrodynamic-type system from (\ref{syst}), i.e.%
\begin{equation*}
(v_{(2)})^{3}-(q^{3}+q^{6})(v_{(2)})^{2}
+(q^{3}q^{6}-q^{4}q^{5}-q^{1})v_{(2)}+q^{1}q^{6}-q^{2}q^{5}=0.
\end{equation*}%
Since the three characteristic velocities are distinct, the commuting
hydrodynamic-type systems (\ref{syst}) can be written in diagonal form (\ref%
{riman}), where the Riemann invariants can be found by quadratures (see (\ref%
{roots})), i.e. (cf. (\ref{rimanful}))%
\begin{equation*}
r^{k}=a+\int (v_{(2)}^{k}db+v_{(3)}^{k}dc),\text{ \ }k=1,2,3.
\end{equation*}

\textbf{Remark}: Hydrodynamic-type system (\ref{six}) possesses at least two
additional local conservation laws. Indeed, the existence of three Riemann
invariants (see (\ref{roots})) implies three additional conservation laws%
\begin{equation*}
\partial _{c}v_{(2)}^{k}=\partial _{b}\frac{%
(v_{(2)}^{k})^{2}-q^{3}v_{(2)}^{k}-q^{1}}{q^{5}},\text{ \ }k=1,2,3.
\end{equation*}%
Hence, the roots $v_{(2)}^{k}$ are conservation law densities. However, five
conservation law densities $q^{3},q^{6},v_{(2)}^{1},v_{(2)}^{2}$, $v_{(2)}^{3}$
are related by the linear equation%
\begin{equation}
q^{3}+q^{6}=v_{(2)}^{1}+v_{(2)}^{2}+v_{(2)}^{3}  \label{lina}
\end{equation}%
by virtue of the Vi\`{e}te theorem. This means that just two of the above three
conservation laws are new.

\textbf{Main Result of this Section}: \textit{Upon expressing} $q^{1},q^{2}$
\textit{and} $q^{6}$ \textit{using the Vi\`{e}te theorem via} $%
q^{3},q^{4},q^{5}$ \textit{and} $v_{(2)}^{1},v_{(2)}^{2},v_{(2)}^{3}$,
\textit{the hydrodynamic-type system} (\ref{six}) \textit{can be written in
the local Hamiltonian form} (\textit{cf}. \cite{Nutku})%
\begin{equation*}
s_{c}^{i}=\tilde{g}^{ik}\partial _{b}\frac{\partial H}{\partial s^{k}},
\end{equation*}%
where the flat coordinates are%
\begin{equation*}
s^{1}=v_{(2)}^{1},\text{ \ }s^{2}=v_{(2)}^{2},\text{ \ }s^{3}=v_{(2)}^{3},%
\text{ \ }s^{4}=q^{4},\text{ \ }s^{5}=q^{5},\text{ \ }%
s^{6}=2q^{3}-(v_{(2)}^{1}+v_{(2)}^{2}+v_{(2)}^{3}),
\end{equation*}%
the metric is%
\begin{equation*}
\tilde{g}_{ik}=-\frac{1}{2}\left(
\begin{array}{cccccc}
-1 & 1 & 1 & 0 & 0 & 0 \\
1 & -1 & 1 & 0 & 0 & 0 \\
1 & 1 & -1 & 0 & 0 & 0 \\
0 & 0 & 0 & 0 & 2 & 0 \\
0 & 0 & 0 & 2 & 0 & 0 \\
0 & 0 & 0 & 0 & 0 & 1%
\end{array}%
\right) ,\text{ \ \ \ }\tilde{g}^{ik}=-\left(
\begin{array}{cccccc}
0 & 1 & 1 & 0 & 0 & 0 \\
1 & 0 & 1 & 0 & 0 & 0 \\
1 & 1 & 0 & 0 & 0 & 0 \\
0 & 0 & 0 & 0 & 1 & 0 \\
0 & 0 & 0 & 1 & 0 & 0 \\
0 & 0 & 0 & 0 & 0 & 2%
\end{array}%
\right) ,
\end{equation*}%
the momentum density $P$ is%
\begin{equation*}
q^{1}=\frac{1}{2}\tilde{g}_{ik}s^{i}s^{k}=\frac{1}{4}%
(s^{1}+s^{2}+s^{3})^{2}-(s^{1}s^{2}+s^{1}s^{3}+s^{2}s^{3})-\frac{1}{4}%
(s^{6})^{2}-s^{4}s^{5},
\end{equation*}%
and the Hamiltonian density $H$ is%
\begin{equation*}
q^{2}=\frac{2s^{1}s^{2}s^{3}+(s^{1}+s^{2}+s^{3}-s^{6})q^{1}}{2s^{5}},
\end{equation*}%
while the other conservation law densities $q^{3},q^{4},q^{5},q^{6}$ (recall
(\ref{lina})) are just linear combinations of the flat coordinates $s^{k}$,
i.e.%
\begin{equation*}
q^{3}=\frac{1}{2}(s^{1}+s^{2}+s^{3}+s^{6}),\text{ \ }q^{4}=s^{4},\text{ \ }%
q^{5}=s^{5},\text{ \ }q^{6}=\frac{1}{2}(s^{1}+s^{2}+s^{3}-s^{6}).
\end{equation*}

\section{Centroaffine Geometry}

Thanks to the presence of the shift symmetry operator (i.e. $h_{1}=\lambda h$%
, see the first equation in (\ref{shift})) linear spectral problem (\ref{lax}%
) for normalized oriented associativity equations (see (\ref{unity})) becomes%
\begin{equation*}
h_{ik}=\lambda \overset{N}{\underset{s=2}{\sum }}u_{ik}^{s}h_{s}+\lambda
^{2}u_{ik}^{1}h,\text{\ }i,k=2,...,N.
\end{equation*}

This linear system arises in centroaffine geometry. Following \cite%
{centroaffine}, consider a linear overdetermined system (which is required
to be compatible of rank $N$)%
\begin{equation}
\frac{\partial ^{2}\mathbf{r}}{\partial a^{i}\partial a^{j}}=\lambda \Gamma
_{ij}^{k}\frac{\partial \mathbf{r}}{\partial a^{k}}+\lambda ^{2}g_{ij}%
\mathbf{r},\quad i,j=2,\dots ,N,  \label{Lax}
\end{equation}%
for the $N$-component position vector $\mathbf{r}=\mathbf{r}(a^{2},\dots
,a^{N})$ of a hypersurface $M^{N-1}$ in centroaffine geometry, where $%
\lambda $ is a spectral parameter, $g_{ij}(a^{2},\dots ,a^{N})$ is a
pseudo-Riemannian metric, and $\Gamma _{ij}^{k}(a^{2},\dots ,a^{N})$ are
components of a torsionless affine connection (which in general is not the
Levi-Civita connection of flat metric $g_{ij}$). The conformal class of $%
g_{ij}$ is nothing but the second fundamental form of $M^{N-1}$.

The compatibility conditions for this system have the form (see \cite%
{centroaffine})%
\begin{equation}
\begin{array}{c}
\Gamma _{ij}^{s}g_{sk}\mathbf{=}\Gamma _{jk}^{s}g_{si}\mathbf{=}\Gamma
_{ik}^{s}g_{sj},\text{ \ \ }\partial _{k}g_{ij}\mathbf{=}\partial
_{i}g_{jk}=\partial _{j}g_{ik},\text{ \ \ }\partial _{k}\Gamma
_{ij}^{s}=\partial _{i}\Gamma _{jk}^{s}=\partial _{j}\Gamma _{ik}^{s}, \\
\  \\
\Gamma _{ij}^{s}\Gamma _{sk}^{m}+g_{ij}\delta _{k}^{m}=\Gamma
_{jk}^{s}\Gamma _{si}^{m}\mathbf{+}g_{jk}\delta _{i}^{m}=\Gamma
_{ik}^{s}\Gamma _{sj}^{m}\mathbf{+}g_{ik}\delta _{j}^{m},%
\end{array}
\label{gen}
\end{equation}%
The equations $\partial _{k}g_{ij}\mathbf{=}\partial _{i}g_{jk}=\partial
_{j}g_{ik},\partial _{k}\Gamma _{ij}^{s}=\partial _{i}\Gamma
_{jk}^{s}=\partial _{j}\Gamma _{ik}^{s}$ mean that the components of the
metric tensor (in these coordinates $a^{k}$) are second derivatives of a
single function $g(\mathbf{a})$ with respect to the corresponding
coordinates $a^{k}$ (i.e. $g_{ik}\equiv \partial ^{2}g/\partial
a^{i}\partial a^{k}$), while the components of the affine connection $\Gamma
_{jk}^{i}$ are also second derivatives of a single vector function $\vec{\Gamma%
}(\mathbf{a})$ (i.e. $\Gamma _{ik}^{s}\equiv \partial ^{2}\Gamma
^{s}/\partial a^{i}\partial a^{k}$, where $\Gamma ^{s}$ are components of
the vector function $\vec{\Gamma}(\mathbf{a})$). Following \cite%
{centroaffine}, one can split the affine connection on two parts, i.e.%
\begin{equation*}
\Gamma _{ik}^{s}=\frac{1}{2}g^{sm}g_{mik}+f_{ik}^{s},
\end{equation*}%
where the first block $\frac{1}{2}g^{sm}g_{mik}$ is a Levi-Civita
connection, while the difference of connections $f_{jk}^{i}$ is a
(1,2)-tensor. Then the equations $\Gamma _{ij}^{s}g_{sk}\mathbf{=}\Gamma
_{jk}^{s}g_{si}\mathbf{=}\Gamma _{ik}^{s}g_{sj}$ (see (\ref{gen})) imply
that the tensor $f_{ijk}=f_{ij}^{s}g_{sk}$ is totally symmetric, defining
the centroaffine cubic form $C=f_{ijk}dx^{i}dx^{j}dx^{k}$ of the
hypersurface $M^{N-1}$ which together with the centroaffine metric $%
M=g_{ij}dx^{i}dx^{j}$ (satisfying compatibility conditions (\ref{gen}))
uniquely characterize a hypersurface. The rest of system (\ref{gen}) are
precisely the normalized oriented associativity equations, where $u^{1}\equiv g$
and all other $u^{k}\equiv \Gamma ^{k}$.

\textbf{The inverse construction}: Now consider an additional variable $%
a^{1} $ and assume that the indices $i,j,k$ run from $1$ up to $N$. Since
the vector function $\vec{c}(\mathbf{a})$ has the components determined from (\ref%
{canon}), we can define the quantities $c_{jk}^{i}(a^{1},\dots, a^{N})$%
\begin{equation}
c_{1k}^{i}=\delta _{k}^{i},\quad i,k=1,\dots ,N,\text{ \ \ \ \ }%
c_{jk}^{i}=\Gamma _{jk}^{i},\quad c_{jk}^{1}=g_{ik},\text{ \ }i,j,k=2,\dots
,N  \notag
\end{equation}%
and $h=(\lambda \mathbf{r},\partial \mathbf{r}/\partial a^{1},...,\partial
\mathbf{r}/\partial a^{N})^{T}$. In such a case, linear system (\ref{Lax})
is replaced again by more general linear system (\ref{zaks}).

Note that these formulas bear considerable resemblance with the formulas for
$c_{jk}^{i}$ for the associativity equations (the famous WDVV equation) at
p. 36 in \cite{centroaffine}. In fact, our formulas reduce to those of
Ferapontov if the metric $g_{ij}$ is constant in the coordinates $a^{i}$: $%
g_{ij}=\tilde{\eta}_{ij}=\mathrm{const}$.

\textbf{Examples}: The oriented associativity equations in three-dimensional
case with the unity condition (\ref{unity}) are nothing but a system of
quadratic equations (see (\ref{syst}) and below)%
\begin{equation*}
u_{bb}=v_{bc}w_{bb}-v_{bb}w_{bc}+w_{bc}^{2}-w_{bb}w_{cc},\text{ \ }%
u_{bc}=v_{cc}w_{bb}-v_{bc}w_{bc},\text{\ \ }%
u_{cc}=v_{bc}^{2}-v_{bb}v_{cc}+v_{cc}w_{bc}-v_{bc}w_{cc}.
\end{equation*}%
Three distinguished versions of the WDVV associativity equations can be
singled out by the special choices of the metrics $\bar{g}^{ik}$ (see (\ref%
{reduc})). If $\eta ^{11}=\eta ^{22}=\eta ^{33}=1$ (and all other entries of $%
\eta ^{ik}$ are equal to zero), then (see (\ref{wdvv}) and (\ref{canon}))%
\begin{equation*}
F=\frac{a^{3}}{6}+\frac{b^{2}+c^{2}}{2}a+z(b,c),
\end{equation*}%
where the three above quadratic equations reduce to a single one ($%
v=z_{b},w=z_{c}$ and $u=(b^{2}+c^{2})/2$)%
\begin{equation*}
z_{bbc}^{2}+z_{bcc}^{2}=1+z_{bbb}z_{bcc}+z_{ccc}z_{bbc};
\end{equation*}%
if $\eta ^{11}=\eta ^{23}=\eta ^{32}=1$, then%
\begin{equation*}
F=\frac{a^{3}}{6}+abc+z(b,c),
\end{equation*}%
where the three aforementioned quadratic equations reduce to a single one ($%
u=bc,v=z_{c},w=z_{b}$)%
\begin{equation*}
1=z_{ccc}z_{bbb}-z_{bcc}z_{bbc};
\end{equation*}%
if $\eta ^{13}=\eta ^{22}=\eta ^{31}=1$, then%
\begin{equation*}
F=\frac{1}{2}(a^{2}c+ab^{2})+z(b,c),
\end{equation*}%
where the three corresponding quadratic equations reduce to a single one ($%
u=z_{c},v=z_{b},w=b^{2}/2$)%
\begin{equation*}
z_{ccc}=z_{bbc}^{2}-z_{bbb}z_{bcc}.
\end{equation*}

The first two examples are associated with the hypersurfaces endowed with 
flat centroaffine metrics (see \cite{centroaffine} for details), while the
third example is related to a non-flat centroaffine metric, because $%
g_{bb}=u_{bb},g_{bc}=u_{bc},g_{cc}=u_{cc}$ and all components of the Riemann
curvature tensor do not vanish.

Thus, we established a link (in fact, an equivalence) between the oriented
associativity equations with unity and centroaffine geometry in a general
non-flat case (cf. \cite{centroaffine}).

\section{Inverse Construction}

In the preceding sections we constructed the transformation from symmetry
consistent conjugate curvilinear coordinate nets to the oriented
associativity equations. In this section, we briefly discuss the inverse
transformation.

\begin{itemize}
\item Any solution of oriented associativity equations (\ref{ori}) is
associated with the corresponding hydrodynamic-type systems (\ref{orient}).

\item Suppose that all characteristic velocities $v_{(k)}^{i}(\mathbf{a})$
of each hydrodynamic-type system (\ref{orient}) are distinct, i.e. the
algebraic equations (for each fixed index $k$)%
\begin{equation}
\det \left\vert c_{jk}^{i}-v_{(k)}\delta _{j}^{i}\right\vert =0
\label{korni}
\end{equation}

have just simple roots. In this paper we restrict our consideration to this
semi-simple case only.

\item If an $N$-component hydrodynamic-type system has pairwise distinct
characteristic velocities (see (\ref{korni})), $N$ conservation laws (see (%
\ref{sohr})) and all components of the Haantjes tensor (see \cite{Haan})
vanish, then this hydrodynamic-type system is semi-Hamiltonian (see \cite%
{Tsar}) and can be written in diagonal form (i.e., $N$ Riemann invariants
exist). The Nijenhuis tensor for the hydrodynamic-type system (\ref{hds})
reads (below in this Section $\partial _{k}\equiv \partial /\partial a^{k}$)
\begin{equation*}
\mathbf{N}_{jk}^{i}=v_{j}^{p}\partial _{p}v_{k}^{i}-v_{k}^{p}\partial
_{p}v_{j}^{i}-v_{p}^{i}(\partial _{j}v_{k}^{p}-\partial _{k}v_{j}^{p}),
\end{equation*}%
and the Haantjes tensor has the form
\begin{equation*}
\mathbf{H}%
_{jk}^{i}=N_{pn}^{i}v_{j}^{p}v_{k}^{n}-N_{jn}^{p}v_{p}^{i}v_{k}^{n}-N_{nk}^{p}v_{p}^{i}v_{j}^{n}+N_{jk}^{p}v_{n}^{i}v_{p}^{n}.
\end{equation*}%
For each system (\ref{orient}) with the time variable $t^{s}$ we readily
find that the Nijenhuis tensor reduces to the form (recall that the summation is
over the pairs of oppositely located repeated indices only)
\begin{equation*}
\mathbf{N}_{(s)jk}^{i}=c_{js}^{q}c_{qks}^{i}-c_{ks}^{q}c_{qjs}^{i},
\end{equation*}%
while the Haantjes tensor becomes%
\begin{equation*}
\hspace*{-7mm}
\mathbf{H}%
_{(s)jk}^{i}=c_{js}^{p}c_{ks}^{m}(c_{ps}^{q}c_{qms}^{i}-c_{ms}^{q}c_{qps}^{i})+c_{ps}^{i}c_{ks}^{m}(c_{ms}^{q}c_{qjs}^{p}-c_{js}^{q}c_{qms}^{p})+c_{ps}^{i}c_{js}^{m}(c_{ks}^{q}c_{qms}^{p}-c_{ms}^{q}c_{qks}^{p})+c_{ps}^{i}c_{ns}^{p}(c_{js}^{q}c_{qks}^{n}-c_{ks}^{q}c_{qjs}^{n}).
\end{equation*}%
This expression formally contains eight blocks. However, the fourth and
fifth blocks coincide. Thus, the Haantjes tensor reduces to the six-block
form%
\begin{equation*}
\mathbf{H}%
_{(s)jk}^{i}=c_{js}^{p}c_{ks}^{m}(c_{ps}^{q}c_{qms}^{i}-c_{ms}^{q}c_{qps}^{i})+c_{ps}^{i}c_{ms}^{q}(c_{ks}^{m}c_{qjs}^{p}-c_{js}^{m}c_{qks}^{p})+c_{ps}^{i}c_{ns}^{p}(c_{js}^{q}c_{qks}^{n}-c_{ks}^{q}c_{qjs}^{n}).
\end{equation*}%
Then one can rewrite the Haantjes tensor in the following form%
\begin{equation*}
\mathbf{H}_{(s)jk}^{i}=c_{js}^{p}c_{ks}^{m}\partial
_{s}(c_{qm}^{i}c_{ps}^{q}-c_{qp}^{i}c_{ms}^{q})+c_{ps}^{i}c_{ks}^{m}\partial
_{s}(c_{ms}^{q}c_{qj}^{p}-c_{mj}^{q}c_{qs}^{p})+c_{ps}^{i}c_{js}^{m}\partial
_{s}(c_{qs}^{p}c_{mk}^{q}-c_{qk}^{p}c_{ms}^{q})
\end{equation*}%
\begin{equation*}
+(c_{qm}^{i}c_{js}^{m}-c_{ms}^{i}c_{qj}^{m})c_{ks}^{p}c_{pss}^{q}+(c_{ms}^{i}c_{qk}^{m}-c_{qm}^{i}c_{ks}^{m})c_{js}^{p}c_{pss}^{q}+c_{qs}^{i}(c_{ks}^{m}c_{mj}^{p}-c_{js}^{m}c_{mk}^{p})c_{pss}^{q},
\end{equation*}%
where each bracket vanishes by virtue of oriented associativity equations (%
\ref{ori}). So, indeed, the family of hydrodynamic-type systems (\ref{orient}%
) is semi-Hamiltonian.

\item In such a case (see (\ref{karak})), the Riemann invariants can be
found by quadratures%
\begin{equation}
r^{i}=\overset{N}{\underset{m=1}{\sum }}\int v_{(m)}^{i}(\mathbf{a})da^{m},
\label{rimanful}
\end{equation}%
where the characteristic velocities $v_{(k)}^{i}(\mathbf{a})$ also (see (\ref%
{korni})) satisfy quadratic relations (cf. \cite{Dubr})%
\begin{equation*}
v_{(j)}^{i}v_{(k)}^{i}=\overset{N}{\underset{s=1}{\sum }}%
c_{jk}^{s}v_{(s)}^{i}.
\end{equation*}%
This means that the functions $a^{n}$ (upon inverting the point
transformation (\ref{rimanful})) can be expressed via the Riemann invariants
$r^{k}$.

\item Then following Tsarev's construction (see \cite{Tsar} for details),
one can compute the so-called Lam\`{e} coefficients (the expressions on the r.h.s.
are equal for all values of $s$)%
\begin{equation}
\partial _{k}\ln \bar{H}_{i}=\frac{\partial _{k}v_{(s)}^{i}(\mathbf{a})}{%
v_{(s)}^{k}(\mathbf{a})-v_{(s)}^{i}(\mathbf{a})},\text{ \ }i\neq k.
\label{lame}
\end{equation}

\item Then following Darboux (see \cite{Darboux} for details), one can find
the rotation coefficients%
\begin{equation}
\beta _{ik}=\frac{\partial _{i}\bar{H}_{k}}{\bar{H}_{i}},\text{ \ }i\neq k,
\label{betta}
\end{equation}%
which satisfy\footnote{%
Since the Lam\`{e} coefficients $\bar{H}_{i}$ are determined up to
multiplication by arbitary functions $\mu _{i}(r^{i})$ (see (\ref{lame})),
they should be fixed by the restriction $\delta \beta _{ik}=0$.} system (\ref%
{darboux}), (\ref{delta}) describing symmetry consistent conjugate
curvilinear coordinate nets. Indeed, since the rotation coefficients are the
same for all commuting flows (\ref{orient}), without loss of generality
consider just $N-1$ commuting hydrodynamic-type systems (\ref{reduk}). Then (%
\ref{rimanful}) reduces to the form%
\begin{equation*}
r^{i}=\tilde{a}^{1}+\overset{N}{\underset{m=2}{\sum }}\int \tilde{v}%
_{(m)}^{i}(\mathbf{\tilde{a}})d\tilde{a}^{m},
\end{equation*}%
where $\tilde{v}_{(m)}^{i}(\mathbf{\tilde{a}})$ depend on the variables $%
\tilde{a}^{2},...,\tilde{a}^{N}$ only. Thus, $\delta \tilde{v}_{(m)}^{i}(%
\mathbf{\tilde{a}})=0$, where the shift symmetry operator (see (\ref{dda}))%
\begin{equation*}
\delta =\frac{\partial }{\partial \tilde{a}^{1}}=\overset{N}{\underset{m=1}{%
\sum }}\frac{\partial r^{m}}{\partial \tilde{a}^{1}}\frac{\partial }{%
\partial r^{m}}=\overset{N}{\underset{m=1}{\sum }}\frac{\partial }{\partial
r^{m}}.
\end{equation*}%
Since the Lam\`{e} coefficients $H_{(1)i}$ can be found from (cf. (\ref{lame}%
))%
\begin{equation*}
\partial _{k}\ln H_{(1)i}=\frac{\partial _{k}\tilde{v}_{(s)}^{i}(\mathbf{%
\tilde{a}})}{\tilde{v}_{(s)}^{k}(\mathbf{\tilde{a}})-\tilde{v}_{(s)}^{i}(%
\mathbf{\tilde{a}})},\text{ \ }i\neq k,
\end{equation*}%
we have $\partial _{k}\delta \ln H_{(1)i}=0$, i.e. $\delta \ln H_{(1)i}=\chi
_{i}(r^{i})$, where $\chi _{i}(r^{i})$ are arbitrary functions. Thus, one
can choose $H_{(1)i}=\tilde{\chi}_{i}(r^{i})\tilde{H}_{(1)i}$ such that $%
\chi _{i}(r^{i})=\partial _{i}\ln \tilde{\chi}_{i}(r^{i})$. Then $\delta
\tilde{H}_{(1)i}=0$, i.e. the corresponding rotation coefficients $\beta
_{ik}$ depend on differences of the Riemann invariants, because $\delta
\beta _{ik}=0$, which follows from (\ref{betta}). So, we conclude that any
solution of the oriented associativity equations (\ref{ori}) gives rise to
some solution of system (\ref{darboux}), (\ref{delta}) describing symmetry
consistent conjugate curvilinear coordinate nets.\looseness=-1
\end{itemize}

\section{The Widest Class of Semi-Hamiltonian Hydrodynamic-Type Systems}

This section is devoted to description and integrability of the widest class
of semi-Hamiltonian hydrody\-namic-type systems (\ref{flux}). A general
solution of oriented associativity equations (\ref{ori}) leads (see
Section \textbf{5}) to a general solution of system (\ref{darboux}), (\ref%
{delta}) describing symmetry consistent conjugate curvilinear coordinate
nets as well as to the basic set of solutions $H_{(i)k}$ (see (\ref{c})). In
this section we construct general solutions of linear spectral problems (\ref%
{a}) and (\ref{b}) which are important for various applications in the
theory of semi-Hamiltonian hydrodynamic-type systems, for instance, in the
generalized hodograph method (see \cite{Tsar}).\looseness=-1

Recall a few major formulas from Sections \textbf{2, 3, 4}.

Suppose that the basic set of solutions $H_{(i)k}$ (see (\ref{c})) of the
linear spectral problem (see (\ref{a}) and (\ref{b}))%
\begin{equation}
\delta H_{i}=\lambda H_{i},\text{ \ \ }\partial _{i}H_{k}=\beta _{ik}H_{i},%
\text{ \ \ \ }i\neq k  \label{d}
\end{equation}%
is found for a given set of rotation coefficients $\beta _{ik}$ depending
only on differences of the Riemann invariants $r^{n}$ (see (\ref{delta}))
and satisfying (\ref{darboux}). Then, the basic set of solutions $\psi
_{i}^{(s)}$ of the adjoint linear problem (see (\ref{a}) and (\ref{b})) is
given by (\ref{psih}), where $\bar{g}^{sn}$ is a non-degenerate symmetric
metric which is inverse to (\ref{metrics}).

\textbf{Our goal} is to find a general solution of the above linear spectral
problem as well as a general solution of the adjoint linear spectral problem%
\begin{equation}
\delta \psi _{i}=\lambda \psi _{i},\text{ \ \ }\partial _{i}\psi _{k}=\beta
_{ki}\psi _{i},\text{ \ \ \ }i\neq k,  \label{e}
\end{equation}%
and general solutions of both other linear spectral problems (\ref{zaks})
and (\ref{bik}).

\textbf{Main result}: $N$ \textit{infinite series of solutions} $%
H_{j}^{(s,k)},\psi _{i}^{(n,p)}$ (\textit{see} (\ref{c})) \textit{can be
found in quadratures}.

Consider the expansions in $\lambda$ of solutions for both linear spectral
problems (\ref{d}) and (\ref{e}), i.e. (see (\ref{c}))%
\begin{equation*}
H_{i}=H_{i}^{(0,k)}+\lambda H_{i}^{(1,k)}+\lambda ^{2}H_{i}^{(2,k)}+...,%
\text{ \ \ }\psi _{i}=\psi _{i}^{(0,k)}+\lambda \psi _{i}^{(1,k)}+\lambda
^{2}\psi _{i}^{(2,k)}+...,\text{ \ }k=1,...,N,
\end{equation*}%
as well as the corresponding expansions of solutions of two other linear
spectral problems (\ref{zaks}) and (\ref{bik}), i.e.%
\begin{equation*}
h=\lambda ^{-1}\delta _{1}^{k}+a^{k}+\lambda c^{k}+\lambda
^{2}h^{(2,k)}+\lambda ^{3}h^{(3,k)}+...,\text{ \ \ }b^{s}=\lambda
^{-1}\delta _{k}^{s}+c_{k}^{s}+\lambda b_{(1,k)}^{s}+\lambda
^{2}b_{(2,k)}^{s}+\lambda ^{3}b_{(3,k)}^{s}+...,
\end{equation*}%
where $\partial _{i}h^{(n,k)}=\psi _{i}^{(n,k)}H_{(1)i}$ and $\partial
_{i}b_{(n,k)}^{s}=\psi _{i}^{(s)}H_{i}^{(n,k)}$, while (see Section \textbf{4%
})%
\begin{equation*}
h^{(n,k)}=\overset{N}{\underset{m=1}{\sum }}\psi _{m}^{(n+1,k)}H_{(1)m},%
\text{ \ \ }b_{(n,k)}^{s}=\overset{N}{\underset{m=1}{\sum }}\psi
_{m}^{(s)}H_{m}^{(n+1,k)}.
\end{equation*}%
Instead of complicated integration of infinite sets of recursion relations (%
\ref{c}) (see (\ref{d}) and (\ref{e})), one can easily integrate the other
infinite sets of recursion relations (see (\ref{zaks}) and (\ref{bik})),
\begin{equation}
dh_{i}^{(n+1,k)}=\overset{N}{\underset{m=1}{\sum }}\overset{N}{\underset{s=1}%
{\sum }}c_{im}^{s}h_{s}^{(n,k)}da^{m},\text{ \ }db_{(n+1,k)}^{i}=\overset{N}{%
\underset{m=1}{\sum }}\overset{N}{\underset{s=1}{\sum }}%
c_{ms}^{i}b_{(n,k)}^{s}da^{m},\text{ \ }i=1,...,N,\text{ \ }n=0,1,...
\label{rec}
\end{equation}

The construction of all these four recursion relations includes the
following steps:
\begin{itemize}
\item Since $\psi _{i}^{(1,k)}\equiv \bar{\psi}_{i}^{(k)}=c_{s}^{k}\psi
_{i}^{(s)}$ (see the proof of the theorem in Section \textbf{3}), where $%
\partial_{i}c_{j}^{k}=\psi _{i}^{(k)}H_{(j)i}$, the expressions $c_{s}^{k}$
can be found in quadratures, i.e.%
\begin{equation*}
dc_{s}^{k}=\overset{N}{\underset{m=1}{\sum }}\psi _{m}^{(k)}H_{(s)m}dr^{m}.
\end{equation*}%
Then (recall that $\partial _{i}a^{s}=\psi _{i}^{(s)}H_{(1)i},\partial
_{i}c^{s}=\psi _{i}^{(1,s)}H_{(1)i}$)%
\begin{equation*}
\psi _{i}^{(1,k)}=\overset{N}{\underset{s=1}{\sum }}\psi _{i}^{(s)}\overset{N%
}{\underset{m=1}{\sum }}\int \psi _{m}^{(k)}H_{(s)m}dr^{m},\text{ \ \ }%
dc^{k}=\overset{N}{\underset{p=1}{\sum }}\overset{N}{\underset{s=1}{\sum }}%
\psi _{p}^{(s)}\left( \overset{N}{\underset{m=1}{\sum }}\int \psi
_{m}^{(k)}H_{(s)m}dr^{m}\right) H_{(1)p}dr^{p}.
\end{equation*}%
Thus, we had a basic set of solutions $H_{(i)k}$. Then we reconstructed the
adjoint basic set of solutions $\psi _{m}^{(k)}$. Then we found $N$
conservation law densities $a^{k}$ and $N$ conservation law densities $c^{k}$%
. So, we have found all structure constants $c_{jk}^{i}$ with unity
condition (\ref{unity}).

\item Then step by step (see (\ref{rec})) we can find $N$ infinite series of
higher conservation law densities $h^{(n,k)}$ as well as $N$ infinite series
of higher conservation law fluxes $b_{(n,k)}^{s}$.

\item Since $\partial _{i}h^{(n,k)}=\psi _{i}^{(n,k)}H_{(1)i}$ and $\partial
_{i}b_{(n,k)}^{s}=\psi _{i}^{(s)}H_{i}^{(n,k)}$, we can find higher
solutions $H_{i}^{(n,k)},\psi _{j}^{(n,k)}$.
\end{itemize}

Moreover, we want to find general solutions of linear systems (\ref{a})
without extra conditions (\ref{b}). This is essential for application of the
generalized hodograph method (see \cite{Tsar}). However, just in some very
special cases general solutions of linear systems (\ref{a}) can be found
explicitly. Nevertheless, using the approach suggested in \cite{Tsar}, one
can see that any initial data for the linear system (\ref{a}) can be
approximated by linear combinations infinitely many particular solutions $H_{i}^{(n,k)},\psi
_{j}^{(n,k)}$ with appropriately chosen coefficients $\xi_{n,k}$.

This means that the general solution of an arbitrary semi-Hamiltonian
hydrodynamic-type system (\ref{flux}) whose rotation coefficients depend on
differences of Riemann invariants only can be found using the generalized
hodograph method (see \cite{Tsar}) and has the form%
\begin{equation*}
x\bar{H}_{i}+tH_{i}=\overset{\infty }{\underset{n=0}{\sum }}\overset{N}{%
\underset{k=1}{\sum }}\xi _{n,k}H_{i}^{(n,k)}.
\end{equation*}

\section{Conclusion}

In this paper, we considered three distinguished objects: oriented
associativity equations, symmetry consistent conjugate curvilinear
coordinate nets and semi-Hamiltonian hydrodynamic-type systems whose
rotation coefficients depend on differences of the Riemann invariants only.
We have shown that these objects are closely related, and thus the knowledge about one of them implies the knowledge about the others.

\section{Acknowledgements}

We thank Boris Dubrovin, Eugeni Ferapontov, Boris Konopelchenko, Franco
Magri and Sergey Tsarev for their stimulating and clarifying discussions.

MVP is grateful to Professor Boris Dubrovin for hospitality in SISSA in
Trieste (Italy) where part of this work has been done. MVP's work was
partially supported by the RF Government grant \#2010-220-01-077, ag.
\#11.G34.31.0005, by the grant of Presidium of RAS ``Fundamental Problems 
of Nonlinear Dynamics" and by the RFBR grant 11-01-00197.

The research of AS was supported in part by the Grant Agency of the Czech
Republic (GA \v{C}R) under the project P201/11/0356, and by the Ministry of
Education, Youth and Sport of the Czech Republic (M\v{S}MT \v{C}R) under
RVO:47813059.

\end{document}